\RequirePackage{fix-cm}
\documentclass[smallextended]{svjour3wide} 
\smartqed  
\usepackage{amsmath,amssymb,bm}
\usepackage{mathrsfs,amsbsy,amssymb,latexsym,amsfonts,amsmath,fancyhdr,lastpage}
\usepackage[dvipdfmx]{graphicx}
\usepackage{epsfig,amsfonts,amsbsy}
\usepackage{fancyhdr}
\usepackage{color}
\usepackage{url}

\newcommand{\pder}[2]{\frac{\partial #1}{\partial  #2}}

\newcommand{\bv}[1]{{\boldsymbol #1}}

\newcommand{\kb}{k_{\rm B}}

\newcommand{\Da}{D^{\rm L}}
\newcommand{\Db}{D^{\rm R}}
\newcommand{\Va}{V^{\rm L}}
\newcommand{\Vb}{V^{\rm R}}
\newcommand{\Na}{N^{\rm L}}
\newcommand{\Nb}{N^{\rm R}}
\newcommand{\Ra}{R^{\rm L}}
\newcommand{\Rb}{R^{\rm R}}

\newcommand{\hatNa}{{\hat N}^{\rm L}}
\newcommand{\hatNb}{{\hat N}^{\rm R}}

\journalname{Journal of Statistical Physics}
\begin{document}

\title{Quasi-static decomposition and the Gibbs factorial in small
  thermodynamic systems}

\author{{\small Shin-ichi Sasa\and Ken Hiura \and  Naoko Nakagawa\and Akira Yoshida}}
\authorrunning{Sasa, Hiura, Nakagawa, Yoshida} 

\institute{
S.-i. Sasa \at
              Department of Physics, Kyoto University, Kyoto 606-8502, Japan 
\and
K. Hiura \at
Universal Biology Institute, The University of Tokyo, Tokyo 113-0033, Japan 
\and 
N. Nakagawa, A. Yoshida\at
              Department of Physics, Ibaraki University, Mito 310-8512, Japan             
}

\date{\today}

\maketitle

\begin{abstract}
For small thermodynamic systems in contact with a heat bath,
we determine the free energy by imposing the following two conditions.
First, the quasi-static work in any configuration change is equal to the
free energy difference. Second, the temperature dependence of the free
energy satisfies the Gibbs-Helmholtz relation.  We find that these
prerequisites uniquely lead to the free energy of a classical system
consisting of $N$-interacting identical particles, up to an additive
constant proportional to $N$. The free energy
thus determined contains the Gibbs factorial $N!$ in addition to the phase
space integration of the Gibbs-Boltzmann factor. 
The key step in the derivation is to construct a quasi-static
decomposition of small thermodynamic systems. 
\end{abstract}


\setcounter{tocdepth}{3}


\section{Introduction}
\label{intro}


Suppose that a small thermodynamic system contacts with a heat bath
of temperature $T$. For equilibrium cases, microscopic states
obey a canonical distribution. Dynamical properties of the system
are described by stochastic processes with the detailed balance 
condition or by projected dynamics of the system from the
Hamiltonian dynamics of the total system containing the heat bath.
The transition from an equilibrium state to another equilibrium state
is caused by a time-dependent parameter. Even for small systems without
the thermodynamic limit, thermodynamic concepts such as the 
first law and the second law are formulated together with incorporating
fluctuation properties \cite{Sekimoto,Seifert,Peliti}.



We study $N$-interacting identical particles
whose Hamiltonian is given by $H(\Gamma;{\cal C})$ with a
microscopic coordinate $\Gamma$ and a set of parameters ${\cal C}$
describing
a system configuration. The equilibrium state is given by the canonical
distribution
\begin{align}
\rho_\beta(\Gamma;{\cal C})=\frac{1}{Z(T,{\cal C})}
\exp(-\beta H(\Gamma;{\cal C})),
\label{can}
\end{align}
where $\beta\equiv 1/(\kb T)$ with the Boltzmann constant $\kb$ and
the partition function $Z(T,{\cal C})$ is given by
\begin{align}
Z(T,{\cal C})=\int d\Gamma \exp(-\beta H(\Gamma;{\cal C}))
\label{part}
\end{align}
from the normalization condition $\int d\Gamma \rho_\beta(\Gamma;{\cal C})=1$.
We then define the free energy $F(T,{\cal C})$ of this system  from the
following two prerequisites. First, for any ${\cal C}_0$ and ${\cal C}_1$,
the free energy difference $F(T,{\cal C}_1)-F(T, {\cal C}_0)$ is equal to the
quasi-static work from ${\cal C}_0$ to ${\cal C}_1$. Because the latter is
expressed by the partition function,
the first condition is expressed as 
\begin{align}
F(T,{\cal C}_1)-F(T, {\cal C}_0)
= -\kb T [\log Z(T, {\cal C}_1) - \log Z(T, {\cal C}_0)].
\label{con:1}
\end{align}
See Sec. \ref{der-con1} for the derivation.
The second condition is that
$F(T,{\cal C})$ satisfies the Gibbs-Helmholtz relation:
\begin{align}
E= \pder{ (\beta F)}{\beta},  
\label{HG}
\end{align}
where $E$ is the expectation of $H(\Gamma;{\cal C})$.


As the simplest configuration, we consider the case where 
$N$-interacting identical particles are  confined
in a cuboid container of volume $V$. The partition function and
the free energy of this system are denoted by  $Z(T,V,N)$ and $F(T,V,N)$.
It has been known that (\ref{con:1}) and (\ref{HG}) lead to
\begin{equation}
  F(T,V,N)=-\kb T [\log Z(T,V,N)-\bar \phi(N)]
\label{F-form}
\end{equation}
with a function $\bar \phi(N)$. See Sec. \ref{der-con1} for a review
of the derivation. Then, the main claim of this paper is that we
can derive 
\begin{align}
  \bar \phi(N)= \log N! +c_0 N
\label{gibbs}
\end{align}
from (\ref{con:1}),
where $c_0$ is an arbitrary constant. Substituting (\ref{gibbs})
into (\ref{F-form}), we obtain
\begin{align}
F(T,V,N)= -\kb T \left[ \log \frac{Z(T,V,N)}{N!} -c_0 N \right].
\label{main}
\end{align}


The formula (\ref{main}) has been explained in standard textbooks
of statistical mechanics. If the system is described by quantum
mechanics, the free energy is defined as
\begin{align}
  F\equiv {\rm Tr}[\hat \rho_\beta \hat H]
  +\kb T {\rm Tr}[\hat \rho_\beta \log \hat \rho_\beta ]
  \label{F-def}
\end{align}
using the density matrix $\hat \rho_\beta({\cal C})$ with
a Hamiltonian $\hat H({\cal C})$.
For $N$-interacting identical particles, the classical limit of
$F$ leads to (\ref{main}) with $c_0= 3\log h$, where $h$ is the Planck
constant. However, when $\hat \rho_\beta$ and $\hat H$ in (\ref{F-def}) 
are replaced by $\rho_\beta(\Gamma;{\cal C})$ and $H(\Gamma;{\cal C})$,
(\ref{F-def}) becomes (\ref{main}) without $N!$.
From this fact, one may understand that $N!$,
which we call the {\it Gibbs factorial}, comes from quantum
mechanics. However, this interpretation is not logically correct,
because it might be possible that the definition of $F$ in (\ref{F-def})
is not valid. There are 
several approaches characterizing the Gibbs factorial in classical
systems \cite{Kampen,Jaynes,Warren,Swedsen,Frenkel,Murashita,Yoshida}. 
Our argument stands on a general principle that thermodynamic quantities
are defined by experimentally measurable quantities. That is, whether or
not $N!$ appears in the formula should be determined by thermodynamic
considerations. This viewpoint has been repeatedly explained
in the literature \cite{Kampen,Jaynes}. 


To determine the functional form of $\bar \phi(N)$ based on (\ref{con:1}),
one may conjecture that a quasi-static decomposition process can be achieved
by inserting a separating wall slowly. If the decomposition were possible
without work  as assumed in macroscopic thermodynamics,
(\ref{con:1}) could lead to 
\begin{equation}
F(T, \lambda V, \lambda N)+  F(T, (1-\lambda) V, (1-\lambda) N)
-F(T,V,N)=0
\label{ext-F}
\end{equation}
for any $N$ and $\lambda$ such that
$\lambda N$ is an integer.
The combination of (\ref{F-form})
with (\ref{ext-F}) gives
\begin{equation}
\bar \phi(N)=  N \log N +c_1 N,
\label{nlogn}
\end{equation}
as shown in Sec. \ref{nlogn-der}, where $c_1$ is an arbitrary
constant. Here,
(\ref{nlogn}) is consistent with (\ref{gibbs})  when $o(N)$
terms are ignored in the thermodynamic limit  $N \to \infty$
and $V \to \infty$ with $V/N$ fixed, while
(\ref{nlogn}) is not equal to (\ref{gibbs})
for finite $N$ cases. The correct statement
is that (\ref{ext-F}) is valid only in the thermodynamic limit.
Thus, the Gibbs factorial $N!$ for macroscopic systems
is understood by thermodynamic considerations, but (\ref{gibbs})
has not been derived from (\ref{con:1})
for finite $N$ cases.


The difference between (\ref{gibbs}) and (\ref{nlogn})
has been studied from the viewpoint
of information thermodynamics \cite{Yoshida}.
Indeed, one can quickly insert a separating wall after measuring
the particle number in the left region of the volume $\lambda V$.
Then, the free energy associated with  measurement-and-feedback
appears on the right-hand side of (\ref{ext-F}).
The form of $\bar \phi(N)$, taking account of this contribution,
is found to be (\ref{gibbs}), not (\ref{nlogn}). 
This argument provides the operational foundation of (\ref{main}).
Therefore, the Gibbs factorial $N!$ would be derived only from quasi-static
works if the quasi-static process corresponding to the free energy
obtained by measurement-and-feedback is constructed. 


Concerning the last point, in Ref. \cite{Horowitz}, it is claimed
that such a quasi-static process exists, but the process is
not explicitly shown.
In the present paper, we construct a quasi-static
decomposition process using a  special confining potential.


Here, it should be noted that for a system consisting of
identical particles, 
the symmetry of the Hamiltonian for any permutation
of particle indexes plays an essential role in the calculation of the
quasi-static work. In contrast, when two types 
of particles, A and B, are mixed, the Gibbs factorial becomes
$N_A!N_B!$, where $N_A$ and $N_B$ are the numbers of particles of
type $A$ and $B$, respectively. In addition to the quasi-static decomposition,
the separation of different types of particles is conducted
using a semi-permeable wall under the assumption that 
the semi-permeable wall can be prepared for any type described 
in the Hamiltonian.


The remainder of the paper is organized as follows. 
In Sec. \ref{pre}, we describe the setup of our study.
We then derive (\ref{con:1}), (\ref{F-form}), and (\ref{nlogn})
as a review. In Sec. \ref{derivation}, we derive
the main result (\ref{gibbs}). 
In Sec. \ref{mixture}, we study binary mixtures and
clarify how the Gibbs factorial is modified from simple
systems. In Sec. \ref{remark}, we briefly comment on
the other approaches presented in Refs. \cite{Warren,Swedsen,Frenkel,Murashita}.
We also consider quantum systems from the viewpoint of quasi-static
decomposition.

\section{Preliminaries }\label{pre}

\subsection{Setup}


We study $N$-interacting particles 
confined in a cuboid region $D$  with $|D|=V$, where 
$|D|$ represents the volume of of the region $D$.
Let $(\bv{r}_i, \bv{p}_i) \in \mathbb{R}^6$ be the position and momentum of
$i$-th particle.  The phase space coordinate $\Gamma$ of the system 
is then given by 
\begin{align}
  \Gamma =(\bv{r}_1, \bv{r}_2, \cdots, \bv{r}_{N},
  \bv{p}_1, \bv{p}_2, \cdots, \bv{p}_{N}).
\end{align}
We assume the Hamiltonian of the system as 
\begin{align}
H_0(\Gamma)  =\sum_{i=1}^{N} \frac{|\bv{p}_i|^2}{2m}
+\sum_{i < j} V_{\rm int}( |\bv{r}_i-\bv{r}_j|),
\label{H0}
\end{align}
where $V_{\rm int}$ represents a short-range interaction
between two particles. For simplicity, we assume that there
exists a microscopic length $\xi$ such that
$V_{\rm int}(r)=0$ for $r \ge \xi$. 


To describe confinement of the $N$ particles
into the region $D$, we introduce a wall potential 
\begin{equation}
V_{\rm wall} (\bv{r};D,\kappa)= \frac{\kappa}{2} |d(\bv{r},D)|^2
\label{v-wall}
\end{equation}
with 
\begin{equation}
d(\bv{r},D)= \inf_{\bv{r}' \in D} |\bv{r}-\bv{r}'|,
\end{equation}
which indicates the distance between a point $\bv{r}$ and
the region $D$. The parameter $\kappa$ in (\ref{v-wall}) 
represents the strength of confinement.
We express the total Hamiltonian as
\begin{align}
H(\Gamma;D,\kappa) =H_0(\Gamma)+\sum_{i=1}^{N} V_{\rm wall}(\bv{r}_i;D,\kappa). 
\end{align}


The canonical distribution for this Hamiltonian is given by
\begin{align}
  \rho_\beta(\Gamma;D,\kappa)=\frac{1}{Z(T,V,N;\kappa)}
      \exp(-\beta H(\Gamma;D,\kappa))
\end{align}
with the partition function 
\begin{align}
Z(T,V,N;\kappa)=\int d\Gamma \exp(-\beta H(\Gamma;D,\kappa)),
\end{align}
where only $V$ dependence of $Z$ is explicitly written in 
the $D$ dependence of $Z$. 
In the hard wall limit $\kappa \to \infty$, we have
the standard partition function
\begin{align}
  Z(T,V,N) &= \lim_{\kappa \to \infty} Z(T,V, N;\kappa), \\
           &= \int d \Gamma 
       \exp(-\beta H_0(\Gamma))\prod_{i=1}^{N}\chi(\bv{r}_{i} \in D),
\end{align}    
where $\chi({\cal A})=1$ if ${\cal A}$ is true,
otherwise $\chi({\cal A})=0$.


More generally, $V_{\rm con}(\Gamma;{\cal C})$ denotes a general confining
potential described by a set of parameters ${\cal C}$. We then write 
\begin{align}
H(\Gamma;{\cal C}) =H_0(\Gamma)+V_{\rm con}(\Gamma;{\cal C}).
\label{Htot}
\end{align}
The canonical distribution and the partition function of this system
are given by (\ref{can}) and (\ref{part}). 
Thermodynamic processes are described by time-dependence of ${\cal C}$
that describes a system configuration.
In particular, for any ${\cal C}_0$ and ${\cal C}_1$,
a quasi-static process connecting them  is represented
by a one-parameter continuous family ${\cal C}_\alpha$,  
where $0 \le \alpha \le 1$, which forms a path in the parameter
space. The quasi-static work done in this process is defined by
\begin{align}
 W ({\cal C}_0 \to {\cal C}_1)
= 
\int_0^1 d\alpha \int d \Gamma \rho_\beta (\Gamma;{\cal C}_\alpha)
\pder{H(\Gamma;{\cal C}_\alpha)}{\alpha}.
\label{qs-work0}
\end{align}

\subsection{Derivation of (\ref{con:1}) and (\ref{F-form})}
\label{der-con1}


We substitute (\ref{can}) into the right-hand side of
(\ref{qs-work0}), and then calculate it as
\begin{align}
 W ({\cal C}_0 \to {\cal C}_1)
&= -\kb T \int_0^1 d\alpha  \frac{1}{Z(T;{\cal C}_\alpha)}
\int d\Gamma\pder{}{\alpha} \exp(-\beta H(\Gamma;{\cal C}_\alpha)) \nonumber \\
&= -\kb T \int_0^1 d\alpha \pder{}{\alpha} \log Z(T;{\cal C}_\alpha) \nonumber \\
&= -\kb T [\log Z(T;{\cal C}_1)-\log Z(T;{\cal C}_0)].
\label{qs-work}
\end{align}
From the condition that the free energy difference is equal
to $W({\cal C}_0 \to {\cal C}_1)$, 
we obtain (\ref{con:1}).

As a special case, we consider the system with a confining potential
\begin{align}
 V_{\rm con}(\Gamma;{\cal C})=  \sum_{i=1}^{N} V_{\rm wall}(\bv{r}_i;D).
 \label{D-con}
\end{align}
Let $F(T,V,N;\kappa)$ be the free energy of the system.
The formula (\ref{con:1}) in this case becomes 
\begin{align}
F(T,V_1,N;\kappa)-F(T,V_0,N;\kappa)=
  -\kb T[\log Z(T,V_1,N;\kappa)- \log Z(T,V_0,N;\kappa)]
\label{F-Z0}
\end{align}
for any $V_0$ and $V_1$. Taking the limit $\kappa \to \infty$, we obtain 
\begin{align}
F(T,V_1,N)-F(T,V_0,N)=-\kb T[\log Z(T,V_1,N)-\log Z(T,V_0,N)].
\label{F-Z}
\end{align}
Note that the formula (\ref{con:1}) cannot be directly
applied to the system with $\kappa \to \infty$ because
$\partial H/\partial \alpha$ is not defined for
the case $\kappa \to \infty$. This is the reason why we need to
pass (\ref{F-Z0}) to get (\ref{F-Z}).
Because the relation (\ref{F-Z}) holds for any $V_0$ and $V_1$,
$F(T,V,N)$ takes the form
\begin{equation}
  F(T,V,N)=-\kb T \log Z(T,V,N)
  + \phi (T,N),
\label{work-piston}
\end{equation}
where a functional form of $\phi (T,N)$ is not determined yet.

Furthermore, the expectation of the energy is calculated as
\begin{equation}
E= - \pder{\log Z}{\beta}.
\end{equation}
The condition (\ref{HG}) is then expressed as
\begin{align}
\pder{(\beta F+\log Z)}{\beta}=0.
\label{HG2}  
\end{align}  
This equality with (\ref{work-piston}) becomes
\begin{equation}
\pder{ \beta \phi(T,N)}{\beta}=0,  
\end{equation}
which leads to
\begin{equation}
\phi(T,N)= \kb T \bar\phi(N) 
\label{p2}
\end{equation}
with a function  $\bar\phi(N)$. Substituting (\ref{p2})
into (\ref{work-piston}), we obtain (\ref{F-form}).

\subsection{Derivation of (\ref{nlogn})}
\label{nlogn-der}

Let $H_{\rm id}(\Gamma;D,\kappa)$ be the Hamiltonian of 
$N$-free particles confined in the same region $D$. The system corresponds to
the ideal gas in thermodynamics. The partition function of the system
is denoted by $Z_{\rm id}(T,V,N)$. Here, putting $\alpha$ in front of
$V_{\rm int}$ in (\ref{H0}), we change $\alpha$ from $\alpha=0$ to $\alpha=1$
gradually. We then have a quasi-static
process from $H_{\rm id}(\Gamma;D,\kappa)$ to $H(\Gamma;D,\kappa)$. 
Because the work done in the quasi-static process
is equal to the free energy
difference of the two systems, the relation
\begin{align}
F(T,V,N)-F_{\rm id}(T,V,N)=-\kb T[\log Z(T,V,N)-\log Z_{\rm id}(T,V,N)]  
\label{toid}
\end{align}
holds. 
By substituting (\ref{F-form}) into (\ref{toid}), we obtain
\begin{equation}
  F_{\rm id}(T,V,N)=-\kb T [\log Z_{\rm id}(T,V,N)-\bar \phi(N)].
\label{F-form-id}
\end{equation}
This expression indicates that $\bar \phi(N)$ is independent of
$H_0(\Gamma)$. Now, we can directly calculate $Z_{\rm id}(T,V,N)$ as
\begin{align}
Z_{\rm id}(T,V,N)= \left( \frac{2m\pi}{\beta}\right)^{3N/2} V^N .
\label{zid}
\end{align}
We substitute (\ref{F-form-id}) with (\ref{zid}) into (\ref{ext-F})
for $F_{\rm id}$. We then have 
\begin{align}
  \bar \phi(\lambda N)+\bar \phi((1-\lambda)N)-\bar \phi(N)=
  N[\lambda \log \lambda +(1-\lambda)\log (1-\lambda)].
\end{align}
Because this relation holds for any $N$ and $\lambda$ such
that $\lambda N$ is an integer, we obtain (\ref{nlogn}).

\section{Derivation of the main result (\ref{gibbs})}
\label{derivation}

\subsection{Outline}

The key step in the derivation of (\ref{gibbs}) is to
construct a quasi-static decomposition. We start with
a description of the decomposed system. Let $\Da$ and $\Db$
be two cuboid regions separated by greater than the interaction
length $\xi$. By using a confining potential with a strength $\kappa$,
we attempt to construct a system where $\Na$ and $\Nb$ particles are almost
in $\Da$ and $\Db$, where $\Na+\Nb=N$ and  $|\Da|=\Va$ and
$|\Db|=\Vb=V-\Va$. Let ${\cal C}_1$ be a parameter set
of the confining potential for the decomposed system, and 
${\cal C}_0$ be the confining potential given in (\ref{D-con}). 
The most important step is to find a continuous family of
confining potentials $V_{\rm con}(\Gamma;{\cal C}_\alpha)$ with
$0 \le \alpha \le 1$ starting from (\ref{D-con}). Here, it is
assumed that a particular $\Na$-particles cannot be selected
by controlling a potential without measuring particle positions.
This property is expressed by
\begin{align}
V_{\rm con}(\Gamma_\sigma;{\cal C}_\alpha)=  
V_{\rm con}(\Gamma;{\cal C}_\alpha)
\label{v-con-p-a}
\end{align}
with 
\begin{align}
\Gamma_\sigma
\equiv(\bv{r}_{\sigma(1)}, \bv{r}_{\sigma(2)}, \cdots, \bv{r}_{\sigma(N)},
  \bv{p}_{\sigma(1)}, \bv{p}_{\sigma(2)}, \cdots, \bv{p}_{\sigma(N)})
\end{align}
for any $\Gamma$ and $\sigma \in P$, where $P$ denotes
a set of all permutations of $\{1,2,\cdots, N\}$.
If such a family of potentials $V_{\rm con}(\Gamma;{\cal C}_\alpha)$,
$0 \le \alpha \le 1$, is found, 
(\ref{con:1}) can be applied to the quasi-static decomposition
process represented by this $({\cal C}_\alpha)_{\alpha=0}^1$.
Then, we take the limit $\kappa \to \infty$ for the obtained
expression. The result is
\begin{equation}
  F(T,\Va,\Na)+F(T,\Vb,\Nb) -F(T,V,N)
  = -\kb T \lim_{\kappa \to \infty}
  [\log Z(T, {\cal C}_1) - \log Z(T, {\cal C}_0)],
  \label{Def2}
\end{equation}
where the numbers of particles in the regions $\Da$ and $\Db$
are fixed as $\Na$ and $\Nb$.
Because
\begin{align}
  Z(T, {\cal C}_0)=Z(T,V,N;\kappa),
  \label{z0}
\end{align}
we have only to calculate $\log Z(T, {\cal C}_1)$.
In the next section, by explicitly constructing the quasi-static
decomposition, we calculate 
\begin{equation}
\lim_{\kappa \to \infty}Z(T, {\cal C}_1)
=
\frac{N!}{\Na! \Nb!}Z(T,\Va,\Na)Z(T,\Vb,\Nb).
\label{key-result}
\end{equation}
Substituting (\ref{F-form}), (\ref{z0}), and (\ref{key-result})
into (\ref{Def2}),  we obtain
\begin{align}
\phi(\Na)+\phi(\Nb)-\phi(N)+\log\frac{N!}{\Na! \Nb!}=0.
\end{align}
Because this holds for any $N$, $\Na$, and $\Nb$, we conclude
(\ref{gibbs}).

\subsection{Derivation of (\ref{key-result})}

\begin{figure}[tb]
\centering
\includegraphics[scale=0.4]{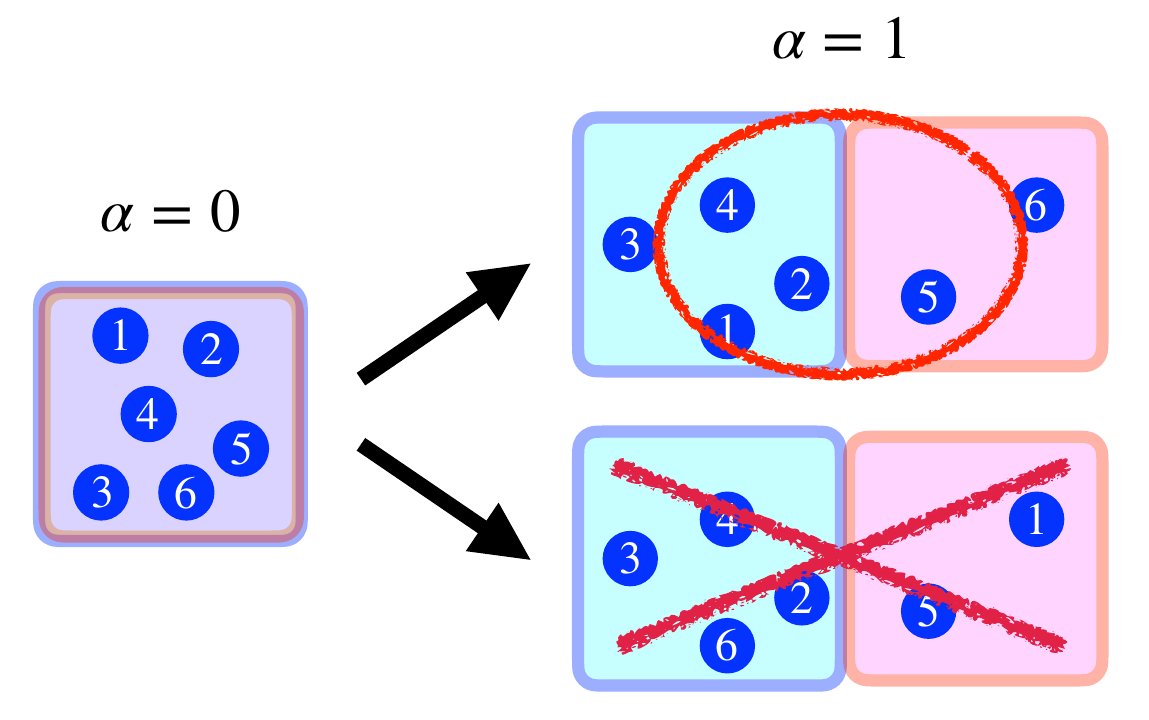}
\caption{
Illustration of a decomposed system using the confining
  potential (\ref{D12-con}). $N=6$, $\Na=4$, and $\Nb=2$.
  Because particles with index $i$
  satisfying $1 \le i \le \Na$ are distinguished from the others,
  this operation is not allowed. }
\label{Fig:qs-decom-0}
\end{figure}

We explicitly construct a confining potential for the decomposed
system described by ${\cal C}_1$, 
where $\Na$ and $\Nb$ particles are almost in $\Da$ and
$\Db$. One may naively consider  such a potential as
\begin{align}
V_{\rm con}(\Gamma;{\cal C}_1)=
\sum_{i=1}^{\Na} V_{\rm wall}(\bv{r}_i;\Da)+
\sum_{i=\Na+1}^{N} V_{\rm wall}(\bv{r}_i;\Db),
\label{D12-con}
\end{align}
illustrated in Fig. \ref{Fig:qs-decom-0}.
The confining potential (\ref{D12-con})
does not satisfy (\ref{v-con-p-a}) with $\alpha=1$. 
This means that  if (\ref{D12-con}) were realized
in a quasi-static process from (\ref{D-con}), 
particles with index $i$ satisfying $1 \le i \le \Na$
could be distinguished from the others. However,
such distinguishment of identical particles 
is not allowed in
the operation. Therefore, we have to look for a potential
that is symmetric with respect to permutations of particle indexes.

As such an example, we propose
\begin{align}
V_{\rm con}(\Gamma;{\cal C}_1)
\equiv \min_{\sigma \in P}
\left[
\sum_{i=1}^{\Na}
V_{\rm wall}(\bv{r}_{\sigma(i)};\Da)
+\sum_{i=\Na+1}^N
V_{\rm wall}(\bv{r}_{\sigma(i)};\Db)
\right] .
\label{v-con}
\end{align}
Because $\Da$ and $\Db$ are separated by greater
than the interaction length $\xi$,
the particle configuration described by a phase space coordinate
$\Gamma$ satisfying $V_{\rm con}(\Gamma;{\cal C}_1)=0$ shows 
that $\Na$ particles are confined in $\Da$ and $\Nb=N-\Na$ particles
are confined in $\Db$, respectively. We here note that particles can
move from one region to the other region when $\kappa$ is finite.
Therefore, all possible decompositions are observed in the
time evolution.

\begin{figure}[tb]
\centering
\includegraphics[scale=0.4]{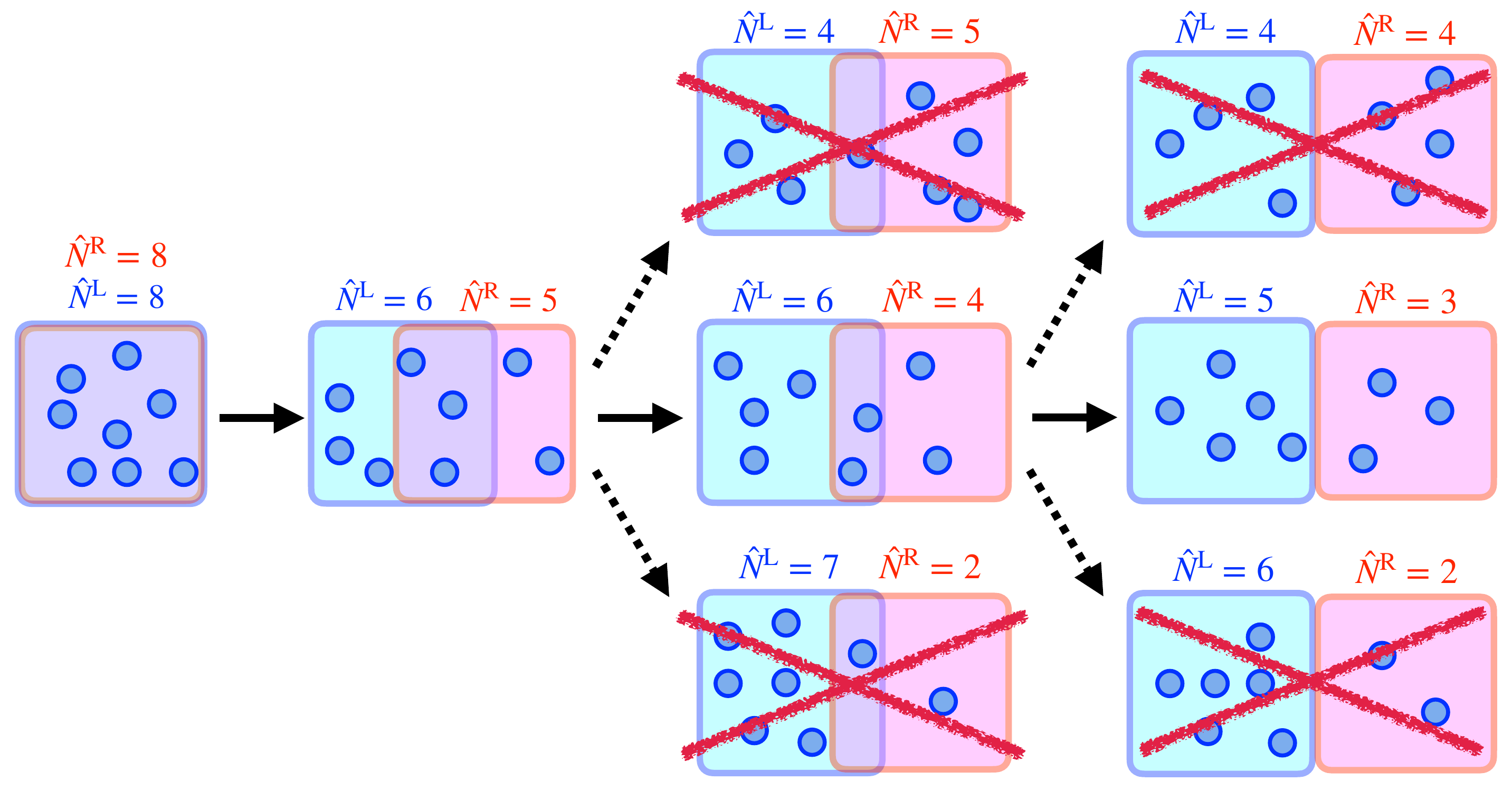}
\caption{
  Illustration of quasi-static decomposition described by
    the family of potentials (\ref{v-con-a}).
  A sequence of
particle configurations. $N=8$, $\Na=5$, and $\Nb=3$.
The case $\alpha=0$ is on the leftmost side, and the case $\alpha=1$
on the rightmost side.
$(\Ra_0,\Rb_0)=(D,D)$, $(\Ra_1,\Rb_1)=(\Da,\Db)$, and
$\Ra_{\alpha} \cap \Rb_{\alpha} \not =\phi, D$ for $0 < \alpha < 1$.  
$\hatNa$ and $\hatNb$ denotes the number of
particles in the region $\Ra_\alpha$ and $\Rb_\alpha$, respectively.
Note that  $\hatNa \ge \Na$ and $\hatNb \ge \Nb$ hold if and
only if $V_{\rm con}(\Gamma;{\cal C}_\alpha)=0$. }
\label{Fig:qs-decom}
\end{figure}

Furthermore, we construct a quasi-static process
from the parameter set ${\cal C}_0$ to ${\cal C}_1$.
Let $\Ra_\alpha$
and $\Rb_\alpha$ ($0\le \alpha \le 1$) be a continuous
family of regions satisfying $(\Ra_0,\Rb_0)=(D,D)$ and 
$(\Ra_1,\Rb_1)=(\Da,\Db)$. 
We then define 
\begin{align}
V_{\rm con}(\Gamma;{\cal C}_\alpha)
\equiv \min_{\sigma \in P}
\left[
\sum_{i=1}^{\Na}
V_{\rm wall}(\bv{r}_{\sigma(i)};\Ra_{\alpha})
+\sum_{i=\Na+1}^N
V_{\rm wall}(\bv{r}_{\sigma(i)};\Rb_{\alpha})
\right].
\label{v-con-a}
\end{align}
To see properties of the potential, we define
  $\hatNa$ and $\hatNb$ as the number of particles
  contained in $\Ra_\alpha$ and $\Rb_\alpha$ for a given 
  particle configuration described by $\Gamma$.
We then confirm that $\hatNa \ge \Na$ and
$\hatNb \ge \Nb$ hold if and only if  
$V_{\rm con}(\Gamma;{\cal C}_\alpha)=0$. Keeping
this in mind, we find that 
(\ref{v-con-a}) with $\alpha=0$
is equivalent to (\ref{v-wall}), and thus (\ref{v-con-a})
with $0 \le \alpha \le 1$ represents the quasi-static
decomposition, as illustrated in Fig. \ref{Fig:qs-decom}.
We also note that
(\ref{v-con-a}) satisfies (\ref{v-con-p-a})
for any permutation $\sigma$.

Now, we calculate $\lim_{\kappa \to \infty}Z(T, {\cal C}_1)$.
For later convenience, we define the subset $G$ of
permutations $P$ as 
\begin{align}
G \equiv \{ \sigma \in P| \sigma(1)< \sigma(2)\cdots <\sigma (\Na);
  \sigma(\Na+1)< \sigma(\Na+2)\cdots <\sigma (N) \}.
\end{align}
For a given phase-space coordinate $\Gamma$ satisfying
$V_{\rm con}(\Gamma;{\cal C}_1)=0$, there is 
a unique $\sigma \in G$ such that 
$\bv{r}_{\sigma(i)} \in \Da$ for $1\le i \le \Na$
and 
$\bv{r}_{\sigma(i)} \in \Db$ for $\Na+1\le i \le N$.
We express this $\sigma$ as $\sigma_\Gamma$ 
to see the $\Gamma$ dependence explicitly. 
We then have
\begin{align}
\lim_{\kappa \to \infty}Z(T, {\cal C}_1)
=\int d\Gamma
\exp(-\beta  H_0(\Gamma))
\prod_{i=1}^{\Na}\chi(\bv{r}_{\sigma_\Gamma(i)} \in \Da)
\prod_{i=\Na+1}^{N}\chi(\bv{r}_{\sigma_\Gamma(i)} \in \Db).
\label{start}
\end{align}
By substituting $\sum_{\sigma \in G}\delta(\sigma, \sigma_\Gamma)=1$
into the right-hand side of (\ref{start}), we rewrite it as
\begin{equation}
\sum_{\sigma \in G} \int d\Gamma
\exp(-\beta  H_0(\Gamma))
\prod_{i=1}^{\Na}\chi(\bv{r}_{\sigma(i)} \in \Da)
\prod_{i=\Na+1}^{N}\chi(\bv{r}_{\sigma(i)} \in \Db).
\label{step}
\end{equation}
Here, using the symmetry property $H_0(\Gamma)=H_0(\Gamma_\sigma)$
for any $\sigma \in P$,
(\ref{step}) is further expressed as
\begin{align}
|G| Z(T,\Va,\Na)Z(T,\Vb,\Nb).
\end{align}
Because 
\begin{align}
|G|=\frac{N!}{\Na! \Nb!},
\end{align}
we obtain (\ref{key-result}).

\section{Free energy of binary mixtures}\label{mixture}

In this section, we study a system consisting of two different types of
particles, A and B, confined in the cuboid region $D$.
Let $\Gamma=(\Gamma^A, \Gamma^B)$
be a phase-space coordinate of the system, where
\begin{align}
  \Gamma^A &
  =(\bv{r}^A_1, \bv{r}^A_2, \cdots, \bv{r}^A_{N^A}
    ,\bv{p}^A_1, \bv{p}^A_2, \cdots, \bv{p}^A_{N^A}), \\
  \Gamma^B &= (
  \bv{r}^B_1, \bv{r}^B_2, \cdots, \bv{r}^B_{N^B},\bv{p}^B_1,
  \bv{p}^B_2, \cdots, \bv{p}^B_{N^B} ).
\end{align}
The Hamiltonian of the system is expressed as
\begin{equation}
H_0(\Gamma)= H^A(\Gamma^A)+H^B(\Gamma^B)+H^{AB}(\Gamma)
\end{equation}
with
\begin{align}
H^A(\Gamma^A) & =\sum_{i=1}^{N^A} \frac{|\bv{p}_i|^2}{2m^A}
+\sum_{i < j} V_{\rm int}^A( |\bv{r}_i^A-\bv{r}_j^A|)
, \\
H^B(\Gamma^B) & =\sum_{i=1}^{N^B} \frac{|\bv{p}_i|^2}{2m^B}
+\sum_{i < j} V_{\rm int}^B( |\bv{r}_i^B-\bv{r}_j^B|)
, \\
H^{AB}(\Gamma) & = \frac{1}{2}
 \sum_{i=1}^{N^A} \sum_{j=1}^{N^B} 
 V_{\rm int}^{AB}(| \bv{r}_i^A-\bv{r}_j^B|). 
\end{align}
Here, $V_{\rm int}^{A/B}$ represents a short-range interaction
between two particles with the same type $A$ or $B$, and
$V_{\rm int}^{\rm AB}$  a short-range interaction between
a particle of type $A$ and a particle of type $B$. The
system configuration that all the particles are confined
in the region $D$ is given by a confining potential 
\begin{align}
V_{\rm con}(\Gamma; {\cal C}_0)
= \sum_{i=1}^{N^A} V_{\rm wall}(\bv{r}_i^A;D)
+\sum_{i=1}^{N^B} V_{\rm wall}(\bv{r}_i^B;D),
\label{semi-con-0}
\end{align}
where ${\cal C}_0$ denotes this set of the parameters. 
The total Hamiltonian
is denoted by $H(\Gamma;{\cal C}_0)$, which is equal to
$H_0(\Gamma)+ V_{\rm con}(\Gamma; {\cal C}_0)$. We then define
the partition function  $Z(T,V,N^A,N^B)$ by
\begin{align}
  Z(T,V,N^A,N^B)=\lim_{\kappa \to \infty}
  \int d\Gamma \exp(-\beta H(\Gamma;{\cal C}_0)).
\end{align}
By a similar argument for the derivation of (\ref{F-form}),
we find  that the free energy $F(T,V,N^A,N^B)$ of the
binary mixture satisfies
\begin{align}
F(T,V,N^A,N^B)=-\kb T [\log Z(T,V,N^A,N^B)-\bar \phi^{(2)}(N^A,N^B)]
\label{v-change}
\end{align}
with a function $\bar \phi^{(2)}(N^A,N^B)$.

In order to determine $\bar \phi^{(2)}(N^A,N^B)$, we consider
a system configuration that particles of type $A$ are in $\Da$,
while particles of type $B$ are in $\Db$, where $\Da$ and $\Db$
are separated by greater than the interaction length of particles.
This configuration is described by a confining potential
\begin{align}
V_{\rm con}(\Gamma; {\cal C}_1)
= \sum_{i=1}^{N^A} V_{\rm wall}(\bv{r}_i^A;\Da)
+\sum_{i=1}^{N^B} V_{\rm wall}(\bv{r}_i^B;\Db),
\label{semi-con-1}
\end{align}
Now, using $\Ra_\alpha$ and $\Rb_\alpha$ in the previous section,
we can construct a quasi-static process parameterized by ${\cal C}_\alpha$
as 
\begin{align}
V_{\rm con}(\Gamma; {\cal C}_\alpha)
= \sum_{i=1}^{N^A} V_{\rm wall}(\bv{r}_i^A;\Ra_\alpha)
+\sum_{i=1}^{N^B} V_{\rm wall}(\bv{r}_i^B;\Rb_\alpha),
\label{semi-con}
\end{align}
which enforces all particles of type A (and B) to be confined
in $\Ra_{\alpha}$ (and $\Rb_\alpha$). This potential corresponds to
a semi-permeable wall for any type of particles. Note that
(\ref{semi-con}) is a standard one-body potential which is
different from (\ref{v-con-a}).
The relation (\ref{con:1}) for the quasi-static process
$({\cal C}_\alpha)_{\alpha=0}^1$ is written as 
\begin{align}
&F(T,V_1,N^A,0)+F(T,V_2,0,N^B)-F(T,V,N^A,N^B)  \nonumber \\
&=-\kb T [\log Z(T,V_1,N^A,0)+\log Z(T,V_1,0,N^B)-\log Z(T,V,N^A,N^B)].
\label{semi-decom}
\end{align}
Here, from the result of the previous section, we have
\begin{align}
  F(T,V_1,N^A,0)&=-\kb T [\log Z(T,V_1,N^A,0)- \log N_A! -c_A N^A],
  \label{g-a} \\
  F(T,V_2,0,N^B)&=-\kb T [\log Z(T,V_2,0,N^B)- \log N_B! -c_B N^B],
  \label{g-b}
\end{align}
where $c_A$ and $c_B$ are arbitrary constants. 
By substituting (\ref{v-change}), (\ref{g-a}), and (\ref{g-b})
into (\ref{semi-decom}), we obtain
\begin{align}
\bar \phi^{(2)}(N^A,N^B)=\log N^A!+\log N^B!+c_A N^A+c_B N^B.
\label{phi2-res}
\end{align}
Therefore, (\ref{v-change}) with (\ref{phi2-res}) becomes
 \begin{align}
   F(T,V,N^A,N^B)=-\kb T \left
   [\log \frac{Z(T,V,N^A,N^B)}{N^A!N^B!}-c_A N^A-c_B N^B \right].
\label{main-bin}
\end{align}

The difference of (\ref{main-bin}) from (\ref{main})
comes from the symmetry of the Hamiltonian for
permutations. The assumption of the argument is that
if the Hamiltonian is not invariant for  the
exchange of two particle indices, the two particles can be
distinguished by the operation of a confining potential.

The above argument implies that the free energy of binary mixtures
can be calculated from works measured in numerical
simulations or experiments. However, from a practical viewpoint,
using the potential (\ref{v-con}) may be impossible in experiments
and is not efficient in numerical simulations. As shown
in Ref. \cite{Yoshida}, the quasi-static decomposition  can be
replaced by the measurement-and-feedback with information
thermodynamics. Furthermore, the expressions using other
quasi-static works can be replaced by the Jarzynski relation
\cite{Jarzynski},
which turns out to be computationally efficient \cite{Yoshida}.

\section{Concluding remarks}
\label{remark}

We have derived (\ref{main}) based on a quasi-static decomposition
using the potential (\ref{v-con}). Our argument stands on the view
that free energy is defined by the measurement of thermodynamic
quantities, that is, whether or not $N!$ appears in the formula should
be determined by thermodynamic considerations. This understanding was
first presented over one hundred years ago \cite{Ehrenfest}.
As explained in Sec. \ref{intro},  it is established that the Gibbs factorial
for macroscopic systems is characterized by quasi-static works.
The achievement in the present paper is that this known result can be
extended to small thermodynamic systems by distinguishing $\log N!$
and $N\log N$. In closing, we briefly discuss other approaches to
understand the Gibbs factorial. 



In Refs. \cite{Warren,Frenkel}, the partition function of a composite
system of sub-systems with $(T,\Va,\Na)$ and $(T,\Vb,\Nb)$
is given as
\begin{align}
  Z_{\rm comp}(T,\Va,\Na,\Vb,\Nb)
  = \frac{N!}{\Na! \Nb!} Z(T,\Va,\Na)Z(T,\Vb,\Nb).
\label{previous}
\end{align}
See Eq. (2) in \cite{Warren} and Eq. (6)  in \cite{Frenkel}.
Although this takes the same form as (\ref{key-result}),
the left-hand side of (\ref{previous}) is not explicitly defined
in Ref. \cite{Warren}. The Hamiltonian of the composite system
should involve a confining potential that is symmetric with
  respect to permutations
of particle indexes. As shown in our paper, we require 
a  special potential (\ref{v-con}), but such an example
is not illustrated in Ref. \cite{Warren}. 
Concerning the last point, in Ref. \cite{Frenkel}, 
the composite system is prepared by making a
small opening in the separating wall of the two sub-systems
with the condition $\Na/\Nb=\Va/\Vb$. However, in this case,
$\Na$ fluctuates, and thus $\Na$ is not a fixed parameter.
Nevertheless, (\ref{previous}) may be justified when the
thermodynamic limit is taken, because $\Na/N$ takes the most probable
value. Note that this argument cannot be applied to small
thermodynamic systems.


As a slightly different approach, in Ref. \cite{Swedsen},
the thermodynamic function is defined by the probability of
unconstrained thermodynamic variables. For example, in isolated
systems, the probability is expressed in terms of  thermodynamic entropy 
\cite{Callen,Einstein}.
For non-interacting particles in contact with a heat bath, the probability
of the number of particles $\Na$ in a given region $\Da$ of the system
is expressed as
\begin{align}
  P(\Na;T,V,N)= \frac{1}{\tilde Z(T,V,N)}
  \exp(-\beta [F(T,\Va,\Na)+ F(T,V-\Va,N-\Na)]),
\label{F-LD}
\end{align}
where $\tilde Z$ is the normalization constant independent of $\Na$.
Now, one may adopt (\ref{F-LD}) as a prerequisite for
determining the free energy. 
Because $P(\Na;T,V,N)$ is calculated from the canonical distribution,
we find that (\ref{main}) holds. The argument is easily generalized to
$N$-interacting identical particles for macroscopic systems, which
is the standard large deviation theory of thermodynamic systems \cite{LD}.
Nevertheless, this approach cannot be used for small thermodynamic
systems except for non-interacting particles.


As a generalization of (\ref{con:1}), the Jarzynski relation
can be used to determine the free energy from non-equilibrium works
\cite{Jarzynski}.
Then, when
a non-equilibrium process without the corresponding time-reversed path
is considered, the relation takes a modified form \cite{MFU}.
In Ref. \cite{Murashita}, the free energy is argued using this
modified Jarzynski relation with absolute irreversibility.
Although the formal result of Eq. (6) in Ref. \cite{Murashita}
is correct, the argument leading to the main result is limited
to non-interacting particles. To prove  their result for
general cases, it is necessary to employ the quasi-static
decomposition introduced in the present paper. We thus believe that
the essence of the Gibbs factorial is in the quasi-static
decomposition, not in the absolute irreversibility. 



Finally, we consider quantum systems. Let $\hat H({\cal C})$
be a Hamiltonian for $N$-interacting identical
particles in a system configuration described by ${\cal C}$.
We define the partition function
$Z_{\rm QM}({\cal C})$ as
\begin{align}
Z_{\rm QM}(T,{\cal C})= {\rm Tr} (\exp(-\beta \hat H({\cal C})).
\end{align}
By repeating a similar argument in Sec. \ref{derivation},
we obtain
\begin{equation}
\lim_{\kappa \to \infty}Z_{\rm QM}(T, {\cal C}_1)
=
Z_{\rm QM}(T,\Va,\Na)Z_{\rm QM}(T,\Vb,\Nb)
\label{key-result-qm}
\end{equation}
instead of (\ref{key-result}). This result leads to
\begin{align}
F(T,V,N)= -\kb T \left[ \log {Z_{\rm QM}(T,V,N)} -c_0 N \right],
\label{main-qm}
\end{align}
instead of (\ref{main}). The free energy $F$ defined by (\ref{F-def})
is equivalent to (\ref{main-qm}) with $c_0=0$. In our viewpoint,
(\ref{F-def}) should not be a starting point, but interpreted as
the result from the prerequisites of free energy. Then, the
free energy for both classical and quantum systems can be uniquely
determined from the quasi-static works without any confusion,
up to an additive constant proportional to $N$. 
The difference between (\ref{main}) and (\ref{main-qm}) comes
from the difference in distinguishability of  microscopic
states for $N$-identical particles. Although this is correct,
the distinguishability of microscopic states
plays no role in the argument determining the free energy.
Rather, the most important issue here is whether or not particles
are distinguished by operation of a confining potential.


\section*{Acknowledgment}

This work was supported by JSPS KAKENHI Grant Numbers 
JP19K03647, JP20K20425, JP22H01144, JP22J00337, and by JST, the
establishment of university fellowships towards the creation of
science technology innovation, Grant Number JPMJFS2105.


\end{document}